\documentstyle[sprocl,epsf]{article}

\newcommand{\mettips}[8]
{
\noindent
\begin{figure}[t]
   \begin{center}
      \leavevmode
	\epsfxsize=11 truecm
       \epsffile{#8}
   \end{center}
    \caption[\protect\small \it #6]{\protect\small \it  #7
      \label{#1}}
\end{figure}
}

\def\Journal#1#2#3#4{{#1} {\bf #2}, #3 (#4)}

\def\PLB{{\em Phys. Lett.}  B}

\def\PRD{{\em Phys. Rev.} D}

\def\mco{\multicolumn}

\def\ra{\rightarrow}

\def\ko{K^0}

\def\be{\begin{equation}}
\def\ee{\end{equation}}
\def\bea{\begin{eqnarray}}
\def\eea{\end{eqnarray}}

\begin{document}
\title{FORCE EVALUATION IN PARTICLE METHODS FOR SELF--GRAVITATING
MULTI--PHASE SYSTEMS}
\author{R. Capuzzo--Dolcetta} \address{Ist. Astronomico, Univ. ``La
Sapienza'', via G.M. Lancisi, 29, I-00161, Rome, Italy}
\author{R. Di Lisio}
\address{Dip. Matematico, Univ. ``La Sapienza'',
P.le Aldo Moro 5, I-00185, Rome, Italy}
\author{P. Miocchi}
\address{Dip. Fisico, Univ. ``La Sapienza'',
P.le Aldo Moro 5, I-00185, Rome, Italy}
\maketitle\abstracts{
A modern approach to the evolution of a mixed (stars and gas) self--gravitating
system is the fully Lagrangian {\it particle} approach. The gaseous 
(particle) phase
differs from the compact because the mutual force is given by the sum
of gravity and pressure gradient. In this note we report of some 
characteristics, advantages and limitations of this approach for what regards
the evaluation of forces in the system. In particular, a comparison between
classic tree--code and fast multipole algorithm to evaluate gravitational
forces is discussed.}

\section{Introduction}

To represent a fluid with a particle method one can consider it
as an ensemble of (smooth) particles, anyone being representative
of a piece of fluid. In this scheme,
each particle is mathematically characterized by a regular ``kernel"
which carries information on the average 
values of dynamical and thermodynamical quantities, as well as on their 
gradients. 
Each particle moves in the force field generated by the whole particle
system, while the associated quantities evolve under their suitably 
regularized laws. We refer the reader to \cite{Monaghan} for a more detailed
description of this method.
In this scheme the difference between the collisional phase (gas) and
the non-collisional one (stars) is that the former pseudo-particles
interact via gravity and pressure, the latter via gravity only.
Of course, one of the most important requirement for the method to work well
is providing a good evaluation of the force field.

\subsection{The ``particle" fluid equations}

In the smooth particle hydrodynamics (SPH) scheme the Lagrangian system of
PDE governing the evolution of a self-gravitating fluid reduces to the set
of ODE:
\begin{eqnarray}
\rho_i &=&\sum_{j=1}^{N}m_j \phi_{h_j}({\bf r}_i-{\bf r}_j),\\
\dot{\bf r}_i &=& {\bf v}_i ,\\
\dot {\bf v}_i &=&
- \sum_{j=1}^{N} m_j \left ({ P_i / \rho^2_i +
 P_j / \rho^2_j  }\right ) \nabla\phi_{h_j}
({\bf r}_i-{\bf r}_j) 
+\nabla U_i, \label{pressure} \\
\dot u_i &=&
\sum_{j=1}^{N}   m_j {P_i / \rho^2_i }
 ({\bf v}_i-{\bf v}_j) \cdot \nabla\phi_{h_j}
({\bf r}_i-{\bf r}_j), \\
\nabla  U_i  &=&  G \sum_{j=1}^{N} {m_j({\bf r}_i-{\bf r}_j) \over |{\bf
r}_i-{\bf r}_j|^3}
\int_0^{|{\bf r}_i-{\bf r}_j|} \phi_{h_j}(s) 4 \pi s^2 ds, \\
f(P_i,\rho_i,u_i)&=&0.
\end{eqnarray}
where the subscript labels the particles, $m_i$ is the
$i$-th particle mass and ${\bf r}_i$ its position, with $i=1,\cdots,N$.
$\phi_{h_i}$ denotes the symmetric kernel associated to the
$i$-th particle whose size
is $h_i$. The other symbols have the usual meaning. In this scheme the
compact non-collisional phase is represented by particles with very small
size, governed by pressureless Eq.s 3 and 4.

Let us fix our attention on the rhs of the motion Eq. 3
given by the sum of:

\noindent
{(i)} $\nabla U$: body (gravity) force, i.e. a
long--range (large scale) force;

\noindent
{(ii)} $\nabla P/\rho$: surface (pressure) force, i.e. a short--range
(small scale) force.

The discretization of the density field with a set of particles
introduces unavoidable
local numerical fluctuations. These fluctuations acts on the evaluation of
terms (i) and (ii) in two very different ways.
The $\nabla P$ term (case (ii))
is, mathematically, a {\it differentiation} of a field
and so it
strongly depends on the short range variations of $P({\bf r})$ (and
on the density via the equation of state).
On the contrary in the term (i) fluctuations
tend to compensate themselves because gravitation
acts on a long range. From a mathematical point of view this can be
understood recalling that the gravitational field is given by a weigthed
{\it integration} on the density distribution
and we know the integration {\it smooths out} the short range variation
of the integrand.

So, the term (i) is very well approximated by particle methods, but this approximation
requires a
amount of computations $\propto N^2$ (as
we will see), while
term (ii) is often hard to handle because of fluctuations but has the
advantage to require
low computation weight.

\section{Pressure force evaluation}

As we said, the pressure term is the most delicate to handle.
We discussed in \cite{dildol} the problem to give a good particle approximation
of this term.
In that paper we studied the error of the SPH evaluation
of the pressure-gradient field
for the set of polytropic compressible fluids in spherical simmetry.
In particular we have studied the error splitting it into a ``modulus" and
a ``directional" part.

We have found that the direction of the pressure-gradient field is well fitted
with large kernels while the best approximation for its modulus is attained
with
kernels such that  every particle interacts eefectively with an almost
constant number
of neighbours, independently of the particle position.
This is very interesting because
allows us to give a general rule for the choice of the kernel size
(in our simulations this number is found to be $0.1 N$).
A good (giving an error less
than $10 \% $) direction of the pressure field
is obtained simply doubling that kernel size.
Unfortunately, this double--kernel scheme requires a CPU time which is
almost 8 times that
required by the
single-kernel evaluation. 
A good balance between the 
quality of the approximation and the CPU weight is reached
fixing the `direction' kernel to be $\root 3 \of 2$ times the
`modulus' one.

In conclusion, a good prescription to estimate the pressure field is
to evaluate the absolute value of the field with a kernel size giving
a constant number of neighbouring particles, while the direction
is estimated using another kernel size as described before.

The CPU time required by this
procedure is less than two times that required by the usual
(single-kernel) particle
evaluation of the pressure field. Thus, this recipe is convenient when
a parallelized code is used.

\section{Gravitational
force evaluation}

As stated in the Introduction, the problem is to
limit the CPU--time without an
appreciable loss of precision.

The so called
{\it particle--particle} calculation consists, simply, in the direct
summation
of the force contributed on every particle 
by all the other particles of the system.
In this way the accuracy is
only limited by the internal round--off error
of the computer but the speed is minumum, being necessary to sum over
$N(N-1)/2\simeq N^2$ pairs.
So the aim of both the methods we tested in \cite{miodol} and \cite{cineca},
namely
the {\it tree--code} and the {\it Fast Multipole Algorithm} (FMA), is to
reduce the number of terms involved in the summation.

\subsection{The comparison between tree-code and FMA}

Detailed descriptions of {\it tree--code} and FMA can be found in
\cite {BH}, \cite {H} and \cite{greengard} respectively.
Their major feature is to reduce the amount of computation to $\propto
N\log N$ via a multipole expansion of the potential. Moreover, FMA
uses also Taylor expansions of the potential.

Performance comparison of the two methods in
astrophysically realistic situations are needed to choose which of them
is worth to be parallelized,
with the aim to radically improve the
time--space resolution of the simulations (see for this \cite{miodol} and
\cite{cineca}).
To do this comparison we built our own optimized serial versions of the codes, that are
slightly different from the ``original'' ones to make them more efficient in
non--uniform situations where adaptivity is important.
For our {\it tree--code} the detailed description can be found in \cite{tesi},
while for the FMA see \cite{miodol}.

The tests of comparison ran on an IBM Risc 6000 workstation
using two different distributions of $N$ particles. In
the first corresponds to a uniform, random distribution of
particles of equal masses.
In the second case, we have distributed the set of particles in such a way to
obtain a discretized profile of the ``clumped'' matter distribution known as
Schuster's  \cite{Schus} profile.

We have considered up to the {\it quadrupole term} in the {\it tree-code},
and up to the {\it second order term} in the multipole expansion in
the FMA (the level of accuracy is in both cases 1 \%).
The tests did not consider the time evolution of the system:
we are only interested to
estimate the CPU--time spent in the evaluation of forces.

\mettips {fig.time}
{}{}{}{}{CPU-time}{CPU-time vs the number of particles in both cases}{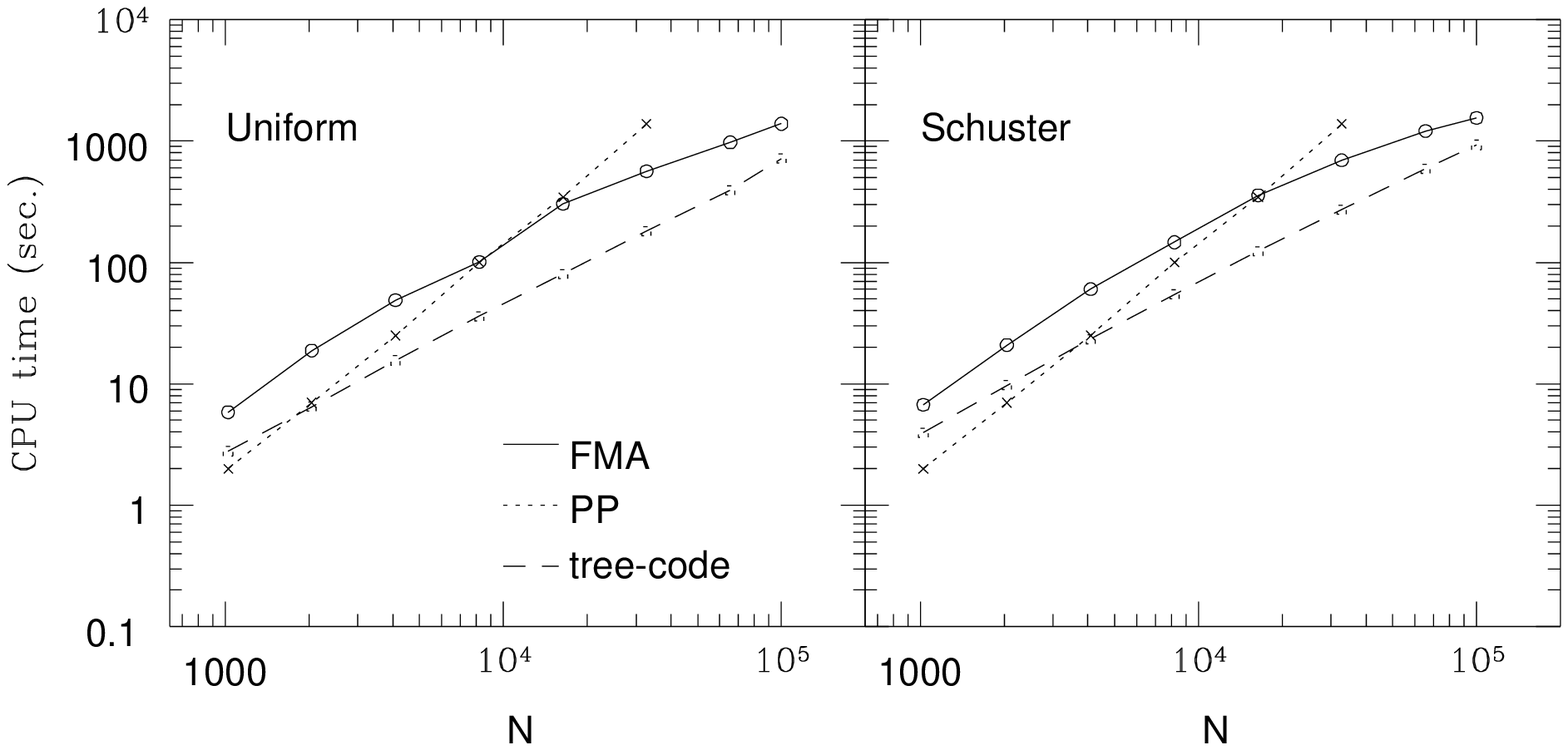}

Figure \ref{fig.time}
gives
the CPU--time nedeed to calculate the accelerations vs. the number $N$ of
particles
(the CPU--time of the {\it particle--particle} method is given for comparison).
Both the algorithms are slower to compute forces in the non--uniform
case than in the uniform one.
This is obviously due to the
more complicated and non-uniform spatial subdivision in clusters of
particles.
In both cases, and for $N$ varying in the range
we tested, the tree--code is faster than the FMA.
So, at least for $N$ in the range
of our tests, the FMA {\it is less efficient than the tree--code} and the
theoretical expectation that its CPU-time were
linear in $N$ is not confirmed by our results.

To conclude we can say that the efforts for the parallelization
have to be done mainly on tree--codes, that appear to be more efficient
than the FMA in single--processor runs. This does not mean that FMA is not
an efficient algorithm. Infact, being easy adding high order multipole term,
it is able to reach accuracies higher than classical {\it tree--codes}, 
which have been conceived to work up to second order term.
Anyway higher accuracy is out of the purposes of
astrophysical simulations where, rather, it is important to deal adequately with
a large range of {\it dynamical time scales}.

\end{document}